\documentclass[reprint,superscriptaddress,showpacs,preprintnumbers,amsmath,amssymb,showkeys]{revtex4-1}
\usepackage{graphicx}
\usepackage{amsmath}

\usepackage{graphicx}
\usepackage{dcolumn}
\usepackage{bm}
\usepackage{graphics}
\usepackage[final]{hyperref}
\usepackage{amssymb}
\usepackage{textcomp}

\begin{document}
\preprint{published in Phys. Rev. A 81, 031604(R) (2010). \href{http://dx.doi.org/10.1103/PhysRevA.81.031604}{DOI: 10.1103/PhysRevA.81.031604}}
\title{Influence of conformational molecular dynamics on matter wave interferometry}

\author{Michael Gring}
\author{Stefan Gerlich, Sandra Eibenberger, Stefan Nimmrichter, Tarik Berrada}
\author{Markus Arndt}%
\homepage{http://www.quantumnano.at} \email{markus.arndt@univie.ac.at} \affiliation{Faculty of Physics, University of Vienna, Boltzmanngasse 5,
A-1090 Wien}
\author{Hendrik Ulbricht}
 \affiliation{School of Physics and Astronomy, University of Southampton, Highfield Southampton SO17 1BJ, United Kingdom}
\author{Klaus Hornberger}
\affiliation{MPI for the Physics of Complex Systems, N\"{o}thnitzer Stra{\ss}e 38, 01187 Dresden}
\author{Marcel M\"uri}
\author{Marcel Mayor}
\altaffiliation{Also: KIT Karlsruhe, Institute for Nanotechnology, P. O. Box 3640, D-76021 Karlsruhe}
\address{University of Basel, Department of Chemistry,
St. Johannsring 19, CH-4056 Basel}
\author{Marcus B\"ockmann}
\address{Lehrstuhl f\"ur Theoretische Chemie, Ruhr-Universit\"at Bochum, D-44780 Bochum, Germany}
\author{Nikos Doltsinis}
\address{Department of Physics, King's College London, London WC2R 2LS, UK}

\begin{abstract}
We investigate the influence of thermally activated internal molecular dynamics on the phase shifts of matter waves inside a molecule interferometer. While de Broglie physics generally describes only the center-of-mass motion of a quantum object, our experiment demonstrates that the {\em translational} quantum phase is sensitive to dynamic conformational state changes {\em inside} the diffracted molecules. The structural flexibility of
tailor-made hot organic particles is sufficient to admit a mixture of strongly fluctuating dipole moments. These modify the electric susceptibility and through this the quantum interference pattern in the presence of an external electric field. Detailed molecular dynamics simulations combined with density functional theory allow us to quantify the time-dependent structural reconfigurations and to predict the ensemble-averaged square of the dipole moment which is found to be in good agreement with the interferometric result. The experiment thus opens a new perspective on matter wave interferometry as it demonstrates for the first time that it is possible to collect structural information about molecules even if they are delocalized over more than hundred times their own diameter.
\end{abstract}


\maketitle

Molecule interferometry is a natural extension of earlier coherence experiments with electrons~\cite{Hasselbach2010},
neutrons~\cite{Rauch2000} or atoms ~\cite{Estermann1930,Cronin2009}. A quest for potential limits of the quantum superposition principle~\cite{Arndt2009b} has recently led to the development of near-field interferometers for complex molecules~\cite{Clauser1994a,Brezger2002,Gerlich2007}.
The term {\em de Broglie} interference usually describes the physics of the center-of-mass motion of a quantum particle. This is why earlier work often emphasized the need for an effective decoupling of the internal states from the external motion~\cite{Kokorowski2001,Hackermueller2004}.

The phase of the translational wavefunction can, however, also be influenced by the interaction between the particle's electromagnetic properties
and the environment: Matter waves were successfully used for characterizing the van der Waals forces in the diffraction of
atoms~\cite{Bruehl2002,Perreault2005} and molecules~\cite{Nairz2003} at nanofabricated gratings.
The combination of electric beam deflection~\cite{Antoine1999a,Knickelbein2001} with quantum interference~\cite{Brezger2002} also showed a new way to measure, for instance, the scalar static~\cite{Berninger2007} and optical polarizability of large molecules~\cite{Hackermueller2007,Gerlich2008}.

In our present work we demonstrate the influence of the {\em internal configurational dynamics} of floppy long molecules on the interference fringe shift inside a near-field matter-wave interferometer. We investigate, in particular, the relevance of the thermally activated internal dynamics for the de Broglie phase evolution in an inhomogeneous electric field.

The general outline of the experiment is as follows: Perfluoroalkyl-functionalized azobenzenes,
C$_{30}$H$_{12}$F$_{30}$N$_{2}$O$_{4}$, were tailor-made to prepare objects of high mass, high vapor pressure and high structural flexibility.
The purified compound was characterized by nuclear magnetic resonance spectroscopy, mass spectrometry and UV/vis and IR-spectroscopy~\cite{Gerlich2007}. The neutral molecules are then evaporated in an effusive source at $T=470\pm5$\,K which determines the mean velocity, the average internal energy and the molecular folding dynamics in the beam. We use a gravitational filtering scheme~\cite{Nairz2000} to select a near-Gaussian velocity distribution that corresponds to a central de Broglie wavelength within the range of $\lambda_{\mathrm{dB}}=2-3$\,pm.

The particles then pass through a Kapitza-Dirac-Talbot-Lau interferometer which consists of a series of three gratings, G$_{1}$ through G$_{3}$. The first and third mask are vertical arrays of 75\,nm wide slits, etched with a period of about d=266\,nm into the supporting SiN$_{x}$ membrane \footnote{grating fabrication: Dr. Tim Savas, MIT Cambridge}. The membrane has a thickness of only 190\,nm and the two material masks are separated by 210\,mm. The diffracting second grating, G$_{2}$, is realized by a focused standing light wave, i.e. by a retro-reflected laser beam with a wavelength of 532\,nm.
The diffraction grating imposes a spatially periodic phase onto the matter wave, which is proportional to the product of the laser power $P$ and the optical polarizability $\alpha_{\mathrm{opt}}(\omega)$ at the laser frequency $\omega$~\cite{Hornberger2009}. This phase is responsible for the interference effect and the observed fringe contrast.
The laser beam divides the distance between the two SiN$_{x}$ gratings exactly in half and intersects the molecular beam under 90\,degrees with an angular uncertainty of better than 500\,$\mu$rad.

The first grating provides the transverse coherence that is required for near-field interference behind the second grating. The coherent molecular evolution around G$_{2}$ leads to a particle density pattern, i.e. the interferogram, at the location of G$_{3}$. An interferogram is then sampled by scanning the position $x_{3}$ of G$_{3}$ in steps of $30$\,nm across the molecular beam while counting all transmitted particles in a quadrupole mass spectrometer. We fit the resulting signal pattern $S(x_{3})$ by the theoretically expected $S(x_3)= O + A\sin(2\pi(x_3-\Delta x_3)/d)$, where $\Delta x_3$ is the offset of the interference fringe.
We determine the experimental quantum fringe visibility as the ratio $V=A/O$. But even more importantly we can quantify with rather high accuracy the shift of the fringe $\Delta x_3$, which allows us to extract valuable information about parameters characterizing the internal molecular state distribution. However, it is not allowed to probe the particles in any way that would reveal their position, as this would destroy the interference pattern.

We therefore apply a conservative force and expose the molecules to an  external electric field $\vec E (x,y,z)$, which is constant in time but inhomogeneous in space. The potential difference across two neighboring interferometer paths imprints a position dependent phase on the matter wave. This results in a shift of the interference pattern at the third grating which reads $\Delta x_3 \propto (\alpha_{\mathrm{stat}}/m) \cdot (U/v)^2 $, where $m$ and $v$ are the mass and the velocity of the particle. The field is created by a voltage $U$ applied to a pair of electrodes between G$_{1}$ and G$_{2}$ whose shapes are carefully designed to ensure that $(\vec E \vec \nabla) \vec E$ is constant within 1\,\% over the diameter of the molecular beam. Within this region, a particle of electric polarizability $\alpha_{\mathrm{stat}}$ will experience the constant force $\alpha_{\mathrm{stat}} (\vec E \vec \nabla) \vec E$. The quantum fringe shift corresponds quantitatively to the classical beam envelope shift induced by the electric force field. It is, however, important to note that the classical prediction of Moir\'e-type fringes is negligible in the parameter regime of our experiment with nanostructured grating masks. The observations of the present experiment can therefore only be explained by including the full quantum treatment for the molecular center-of-mass motion.

In the past, the described setting was used to determine static molecular properties of rigid molecules, such as the polarizability of fullerenes~\cite{Berninger2007}. The present study tackles a more complex phenomenon since we are dealing with extended, structurally flexible particles whose thermally driven configuration dynamics must be accounted for to understand the center-of-mass phase evolution.

As already seen in earlier classical experiments~\cite{Compagnon2002} thermally activated vibrations may induce dynamic electric dipole moments $\mathbf{d}$ even in molecules that are point-symmetric and non-polar in their thermal ground state.
If the number of accessible internal states is large the total linear response of the molecule to the E-field is no longer described by the polarizability but rather by the van Vleck expression~\cite{Vleck1965,Bonin1997} for the electric susceptibility
\label{vanVleck}
$    \chi = \alpha_{\mathrm{stat}}+ \left \langle \, \mathbf{d}^2 \,\right \rangle/3k_BT = \alpha_{\mathrm{stat}} + \alpha_{\mathrm{dip}}.
$
Here, $\textbf{d}^2$ is thermally averaged over the energy landscape that is accessible to the vibronic motion in the absence of the external field at an internal temperature T. As we operate an effusive beam source it is well justified to assume that this rotational and vibrational temperature equals the source temperature of $470\,\pm5$\,K.

In order to get an estimate for the dependence of the polarizability and dipole moment on the molecular conformation, electronic structure calculations
were performed using density functional theory (DFT). The underlying conformational structures were generated by a molecular dynamics simulation (MDS) using
the GROMOS package~\cite{Gunsteren1996} and a tailor-made force field~\footnote{see supplementary material at \href{http://dx.doi.org/10.1103/PhysRevA.81.031604}{DOI: 10.1103/PhysRevA.81.031604}} based on a recent parametrization for azobenzene
derivatives~\cite{Boeckmann2007}. During the simulation, a single molecule was propagated for a total of 100~ns with a time step of 1~fs
following a 100~ps equilibration period. The simulated temperature was controlled by a Berendsen thermostat with a coupling constant of 0.1~ps. Assuming that the MDS time evolution covers the entire conformational phase space, its trajectory represents the statistical ensemble of conformations in the hot molecular beam. Along the trajectory, the molecular structure was extracted every 5~ns and fed into a DFT (B3LYP/6-31+G*) calculation using
the Gaussian package~\cite{Frisch2003short} to obtain the electronic polarizability and the absolute value of the dipole moment.

\begin{figure}[t]
  \begin{center}
\includegraphics*[width=0.95\columnwidth]{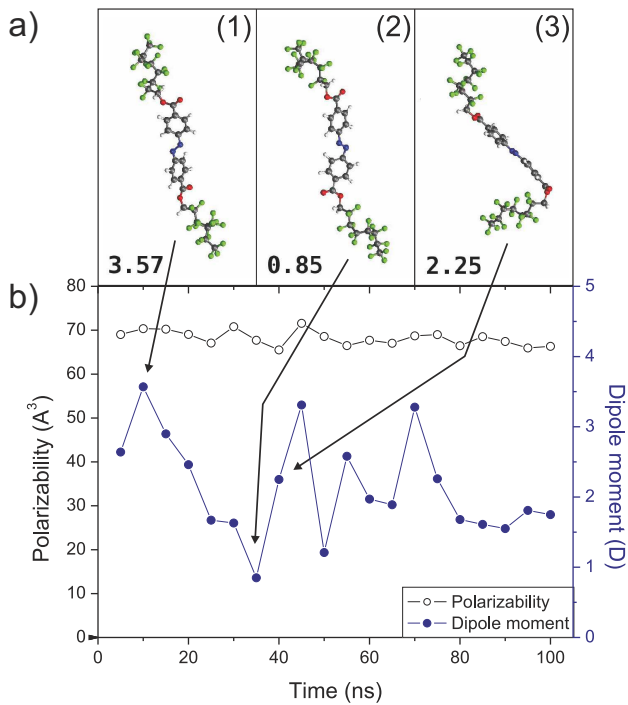}
  \end{center}
  \caption{a) Snap shots of the MDS of perfluoroalkylated azobenzenes
at 10~ns (1), 35~ns (2), and 40~ns (3). One clearly recognizes the varying exposure of different side chains towards external fields. Although
the molecule has no permanent static electric dipole moment in the thermal ground state, excitation at 500\,K is associated with a rapid
evolution and the expression of temporary dipole moments\,(in Debye, D). b) Static scalar polarizability, $\alpha_{\rm stat}$ and electric
dipole moment along the MDS trajectory. While $\alpha_{\rm stat}$ is essentially constant, $d$ varies by as much as 300\,\%.}
\label{fig:snaps}
\end{figure}

Three of the snapshots are depicted in Fig.~\ref{fig:snaps}a together with their total dipole moments. The computed
polarizabilities and moments of 20 molecular structures, taken in time steps of 5\,ns, are displayed in Fig.~\ref{fig:snaps}b. While the
polarizability is only mildly affected by the large conformational changes, with an average value of $\alpha_{\mathrm{stat}}=68.2$~\AA$^3\times
4\pi \varepsilon_0$, the dipole moment fluctuates strongly between $0.8\,$Debye (D) and $3.6\,$D. The most relevant structural changes are due to a rotation of the aromatic rings around the N-C bond and a rotation of the alkyl chain around the C-O axis. The simulated values displayed in Fig.~\ref{fig:snaps}b thus sum up to a thermally averaged polarizability of $\alpha_{\mathrm{dip}}=24.6$~\AA$^3\times 4\pi
\varepsilon_0$ and a total electric susceptibility of $\chi=92.8$~\AA$^3\times 4\pi \varepsilon_0$.

We estimate the uncertainty of the mean values from our computational sample of size $n=20$ using a t-test where we assume normally distributed variables over the whole MDS trajectory. The true thermal averages of $\alpha_{\mathrm{stat}}$ and $\alpha_{\mathrm{dip}}$ are contained in the confidence intervals $\pm 0.8\,$\AA$^3$ and  $\pm 7.6\,$\AA$^3$ around their respective sample averages.
The total confidence interval for the susceptibility is given by $\chi=92.8 \pm 0.8 \pm 7.6$~\AA$^3\times 4\pi \varepsilon_0$ at a statistical significance level of $5\,\%$. The calculation reveals that the molecular susceptibility is substantially larger than the static polarizability alone and that it should be possible to get quantitative information about thermodynamically driven internal molecular dynamics through the van Vleck formula, if both $\chi$ and $\alpha_{\mathrm{stat}}$ can be measured independently.
\begin{figure}[tbp]
   \centering
   \includegraphics[width=0.95\columnwidth]{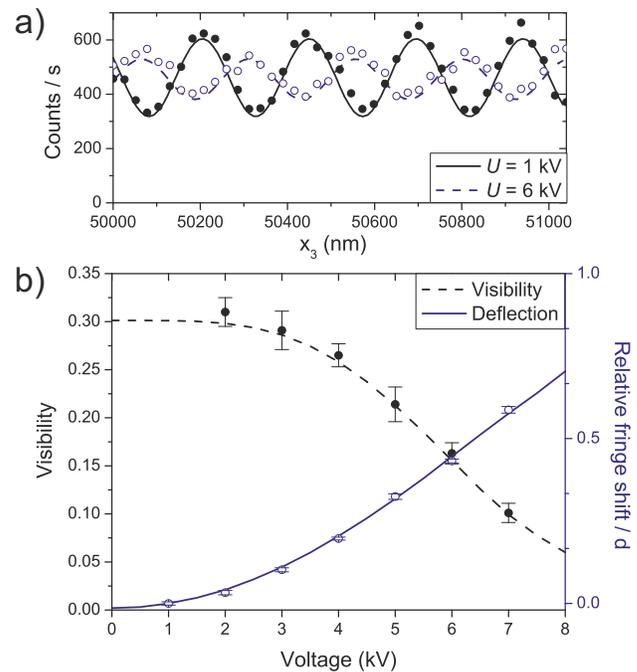}
   \caption{a) Two typical interference patterns at $U=1$kV (full circles) and $U=6$kV (hollow circles). b) Interference fringe shift (right scale) and visibility (left scale) as a function of the deflection voltage. Hollow circles: measured shift. Continuous line: Quantum fit (see text) from which we extract the molecular susceptibility. Full circles: experimental fringe visibility. Dashed line: expected decrease of the fringe visibility for the experimentally determined velocity distribution and susceptibility. Error bars: $1\sigma$-uncertainty of the fit to the raw data.
} \label{Susceptibility}
  \end{figure}

The outcome of the deflection experiment is summarized in the lower panel of Fig.~\ref{Susceptibility} where the mean visibility and the shift of the interference pattern are plotted against the electrode voltage.

After a calibration of the experimental geometry factors using fullerenes~\cite{Berninger2007}, a fit of the expected and measured fringe shift as a function of the electrode voltage allows us to determine the total susceptibility for the functionalized azobenzenes. We find an experimental value of $\chi= 95\pm 3 \pm 8$~\AA$^3\times 4\pi \varepsilon_0$ where the first and second uncertainty value represent the statistical and the systematic error respectively. The experimental result is in good agreement with the MDS value $\chi= 92.8~$\AA$^3\times 4\pi \varepsilon_0$.

It is important to see that the velocity distribution of the selected thermal beam affects the result in two ways: on the one hand, slow molecules
experience a larger phase shift of the fringes than the fast ones. This also explains the slight deviation of the experiment from the quadratic expectation. On the other hand, the dispersiveness of the phase shift also affects the fringe visibility, as shown by the dashed line in the same panel. Both can be accounted for in a quantum calculation based on~\cite{Hornberger2009}, which is in very good agreement with the observation. The velocity distribution in Fig.~\ref{Susceptibility} is centered on
$v_\mathrm{mean}= 146$\,m/s with a width of $\Delta v_{\mathrm{FWHM}}=31$\,m/s.

Our setup now also offers, in addition, a unique way for separately measuring the electronic contribution to $\chi$. In the optical field of the light grating the response of the molecules is no longer influenced by changes in the molecular structure. The nuclear motion is too slow to follow
the rapid field oscillations. All structural dipole contributions will therefore be averaged out.
Our Kapitza-Dirac-Talbot-Lau interferometer thus allows us to determine the optical molecular polarizability $\alpha_\mathrm{opt}(\omega)$ at the grating laser frequency $\omega$~\cite{Gerlich2007}.

In atoms $\alpha_\mathrm{opt}(\omega)$ is strongly frequency-dependent and differs substantially from the static value. However, in large molecules and in particular in those used here, all molecular absorption lines are concentrated in the ultra-violet region. Hence, the static polarizability is well approximated by the optical polarizability $\alpha_{\mathrm{stat}}\simeq \alpha_{\mathrm{opt}}$, as also known, for instance, for the case of fullerenes~\cite{Dresselhaus1998}. We extract $\alpha_{\mathrm{opt}}$ from a fit of the theoretical expectation to the measured $V(P)$-curve, i.e. from a plot of the interference fringe visibility as a function of the laser power $P$.

\begin{figure}[h]
   \centering
   \includegraphics[width=0.95\columnwidth]{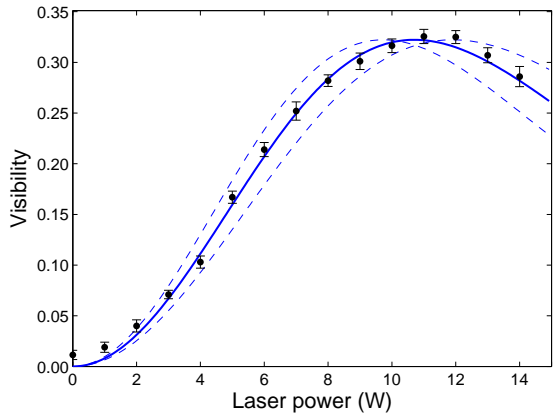}
   \caption{Fringe visibility as a function of the diffracting laser power. Full circles: Average of three consecutive measurements. Error bars: standard deviation. Full line: calculated quantum curve with the optical polarizability $\alpha_{\mathrm{opt}}$ as the only free parameter. Dashed lines: Same fit for $\alpha_{\mathrm{opt}} \pm 10\,\%$. The measured velocity distribution ($v_\mathrm{mean}= 140$\,m/s, $\Delta v_{\mathrm{FWHM}}=28$\,m/s) is included in the simulation of the fringe visibility. } \label{OpticalPolarizability}
\end{figure}

The experimental result is shown as full circles in Fig.~\ref{OpticalPolarizability}. It is well reproduced by a quantum calculation~\cite{Hornberger2009} which contains the optical polarizability $\alpha_{\mathrm{opt}}$ as the only free parameter (solid line). An additional fit that includes a finite optical absorption cross section as a second parameter is consistent with the result $\sigma_{\mathrm{abs}} (532\,\mathrm{nm}) < 2\times 10^{-18}$cm$^2$, and justifies the omission of this factor. This procedure yields a value of $\alpha_{\mathrm{opt}} = 61\,\pm  1 \pm 7 \,\AA^3\times 4\pi \varepsilon_0$. This agrees, again, with the numerical simulation for the static polarizability, shown in Fig.~\ref{fig:snaps}, within the experimental uncertainty.
The uncertainty is composed of a statistical (first) and a systematic contribution (second). The first value represents the standard deviation of
the experimental scatter in three independent experiments, each composed of two or three measurement runs. The systematic part is determined by
possible uncertainties in the overlap between the laser field and the molecular beam, in the velocity distribution and the measurement of
the laser power which is known to within $\pm 10\,\%$. The dashed curves in~Fig.~\ref{OpticalPolarizability} simulate the expected power dependence of the fringe visibility  for optical polarizabilities 10\% above and below the best fit.

Knowing the total susceptibility $\chi$ and the static polarizability $\alpha_{\mathrm{stat}}$ we are now able to identify the presence of a dipolar contribution to the total susceptibility. Its value is in good quantitative agreement with the theoretical prediction.

In conclusion, our work shows for the first time, that even pure de Broglie interference allows us to get access to thermally activated time-averaged internal dynamics of molecules.

Our work has been supported by the Austrian Science Funds FWF within the Wittgenstein program Z149-N16 and the doctoral program W1210 CoQuS as
well as by the ESF EUROQUASAR Program MIME. The synthesis of the molecular compounds has been supported by the Swiss National Science Foundation
(SNSF) and the Innovation Promotion Agency (CTI). MB is funded by the VW Foundation.

%

\end{document}